\documentclass[aps, prd, amsmath, floats, floatfix, twocolumn, nofootinbib]{revtex4-1}
\usepackage{graphicx}
\usepackage{color}

\newcommand{\be}{\begin{eqnarray}}
\newcommand{\ee}{\end{eqnarray}}
\newcommand{\rar}{\rightarrow}

\begin{document}

\title{Testing the Bardeen metric with the black hole candidate in Cygnus X-1}

\author{Cosimo Bambi}
\email{bambi@fudan.edu.cn}

\affiliation{Center for Field Theory and Particle Physics \& Department of Physics, Fudan University, 200433 Shanghai, China}

\date{\today}

\begin{abstract}
In general, it is very difficult to test the Kerr-nature of an astrophysical black hole
candidate, because it is not possible to have independent measurements of both
the spin parameter $a_*$ and possible deviations from the Kerr solution. Non-Kerr 
objects may indeed look like Kerr black holes with different spin. However, 
it is much more difficult to mimic an extremal Kerr black hole. The black 
hole candidate in Cygnus~X-1 has the features of a near extremal Kerr black hole, 
and it is therefore a good object to test the Kerr black hole paradigm. The 
3$\sigma$-bounds $a_* > 0.95$ and $a_* > 0.983$ reported in the literature and 
valid in the Kerr spacetime become, respectively, $a_* > 0.78$ and $|g/M| < 0.41$, 
and $a_* > 0.89$ and $|g/M| < 0.28$ in the Bardeen metric, where $g$ is the 
Bardeen charge of the black hole.
\end{abstract}

\maketitle

%%%%%%%%%%%%%%%%%%%%%%%%%%%%%%%

\begin{figure*}
\begin{center}
\hspace{-0.5cm}
\includegraphics[type=pdf,ext=.pdf,read=.pdf,width=8.5cm]{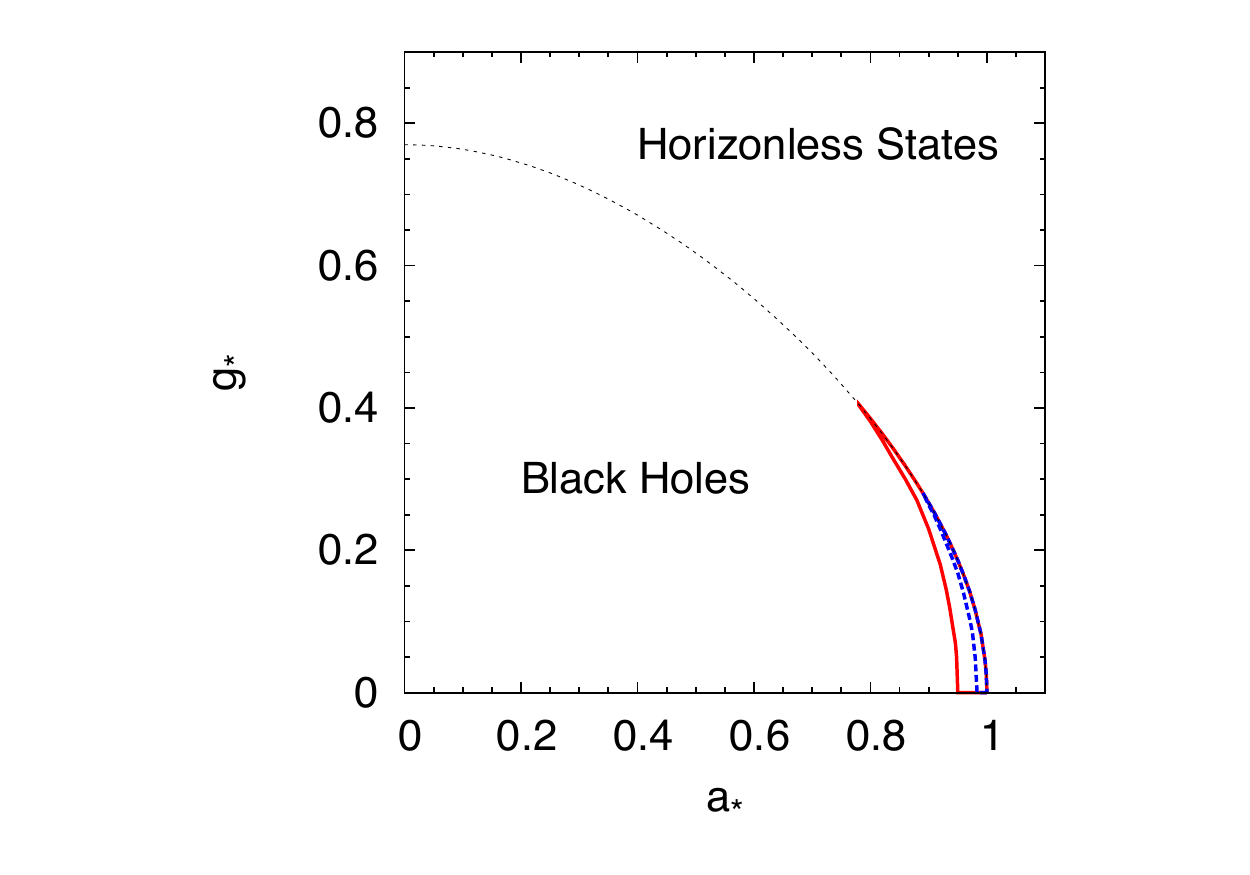}
\includegraphics[type=pdf,ext=.pdf,read=.pdf,width=8.5cm]{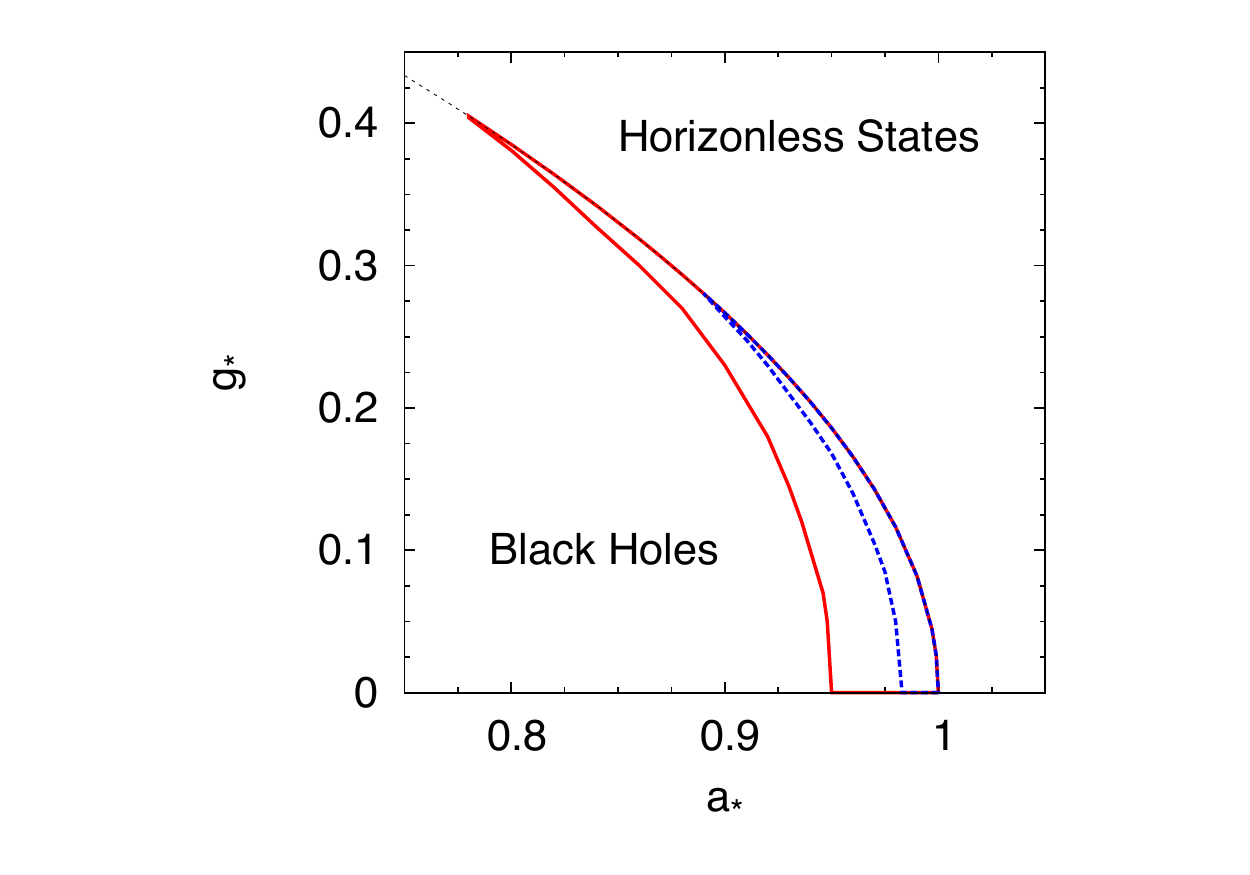}
\end{center}
\vspace{-0.5cm}
\caption{Spin parameter-deformation parameter plane of the Bardeen metric. 
The thin black dashed line separates BHs from configurations without an event 
horizon (i.e. $\Delta = r^2 - 2 m r + a^2 = 0$ has no real roots). The thick red 
solid line and the thick blue dashed line are the boundaries of the allowed 
regions for the BH candidate in Cygnus~X-1 inferred, respectively, from the 
3$\sigma$-bound obtained in the Kerr metric $a_* > 0.95$ in Ref.~\cite{cyg}
and $a_* > 0.983$ in Ref.~\cite{cyg2}. The right panel is an enlargement of the
left panel. See the text for more details.}
\label{fig1}
\end{figure*}

Astrophysical black hole (BH) candidates are supposed to be the Kerr BHs 
predicted in general relativity, but the actual nature of these objects has still 
to be verified~\cite{review}. At present, there are only two relatively robust 
techniques capable of probing the geometry of the space-time around BH 
candidates; that is, the continuum-fitting method~\cite{cfm} and the analysis of 
the K$\alpha$ iron line~\cite{iron}. Under the Kerr BH paradigm, these techniques 
can provide an estimate of the spin parameter $a_* = J/M^2$, where $M$ and 
$J$ are, respectively, the BH mass and spin angular momentum\footnote{Throughout 
the paper, I use units in which $G_{\rm N} = c = 1$, unless stated otherwise.}. Both the 
continuum-fitting method and the iron line analysis have been extended to 
non-Kerr spacetimes to test the nature of BH candidates~\cite{cb,other}.
However, in general it is only possible to constrain some combination between 
the spin and possible deviations from the Kerr solution: the spectrum of 
non-Kerr objects can be very similar to the one of Kerr BHs with a different spin 
parameter.

While a number of very exotic metrics can already be ruled out by current 
observations with these techniques~\cite{tests}, some more theoretically 
motivated non-Kerr metrics are very difficult to test and the combination of the 
continuum-fitting method and the iron line analysis cannot solve the degeneracy 
between the spin and the deformation parameters~\cite{cb2}. The difficulty to 
constrain deviations from the Kerr metric independently of the measurement 
of the spin is common to other (potentially more sophisticated) approaches, 
like the observation of quasi-periodic oscillations~\cite{n1} and of the X-ray 
polarization~\cite{n2}. This problem might be fixed in the future, at least partially, by 
combining one of the techniques above with the estimate of jet power~\cite{n3} or 
the measurement of the BH shadow~\cite{n4}. However, the observational 
features of near extremal Kerr black holes are peculiar, and therefore the 
observation of a BH candidate that looks like a near extremal Kerr BH can 
constrain the nature of these objects.

While the analysis of the iron line profile is potentially more powerful, the 
continuum-fitting method is based on more solid physics, the uncertainty on the
final measurement is more reliable, and therefore this technique is at present 
more suitable to derive conservative but robust bounds. It is based on the 
study of the soft X-ray component of stellar-mass BH candidates, which 
is interpreted as the thermal spectrum of a geometrically thin and optically thick 
accretion disk. Among the BH candidates studied with the continuum-fitting
method, there are two objects that look like near extremal Kerr BHs: 
GRS~1915+105~\cite{grs} and Cygnus~X-1~\cite{cyg,cyg2}. Studies of the 
iron line analysis support this conclusion~\cite{iron2} (but see Ref.~\cite{miller}, 
whose discrepancy is probably due to the improper data state -- not high/soft 
state -- and the improper usage in the continuum model in extracting the
skewed iron line profile~\cite{cyg}).
GRS~1915+105 is a very peculiar source, whose data are difficult to interpret,
and the measurement reported in Ref.~\cite{grs} is based on the assumption
that the jet observed in this source is perpendicular to the accretion disk, 
which is at least questionable. The studies in Refs.~\cite{cyg,cyg2} of Cygnus~X-1
are more recent and the results may be thought to be more robust. Under the
assumption that the geometry of the spacetime around this object is described 
by the Kerr metric, the 3$\sigma$-bound on the spin parameter of the BH
candidate in Cygnus~X-1 is found to be $a_* > 0.95$ in~\cite{cyg} and 
$a_* > 0.983$ in~\cite{cyg2}.

The aim of this letter is to test the Bardeen metric, which describes the spacetime 
of a singularity-free BH and can be formally obtained by coupling Einstein's 
gravity to a non-linear electrodynamics field~\cite{bar0}. In Boyer-Lindquist
coordinates, the metric of the rotating solution is equivalent to the Kerr metric
with the mass $M$ replaced by the function $m$, given by~\cite{bar1}
\be
M \rar m = M \left(\frac{r^2}{r^2 + g^2}\right)^{3/2} \, ,
\ee
where $r$ is the radial coordinate and $g$ is the magnetic charge of the non-linear
electrodynamics field (or simply the deformation measuring the deviations 
from the Kerr metric). As $g$ has the same dimension as $M$, in analogy with
the dimensionless spin parameter $a_*$ it is convenient to introduce the 
dimensionless deformation parameter $g_* = g/M$. Let us note that $m \rar M$ at 
large radii. The radius of the event horizon, which is given by the largest root of 
$\Delta = r^2 - 2 M r + a^2 = 0$ (where $a = J/M$ is the specific spin angular momentum) 
in the Kerr spacetime, is now the largest root of $\Delta = r^2 - 2 m r + a^2 = 0$. Bardeen 
BHs exist only below a critical spin parameter $a_*^{\rm c}$, which depends 
on the value of $g_*$ and reduces to the Kerr one $a_*^{\rm c} = 1$ for $g_* = 0$, 
while for $a_* > a_*^{\rm c}$ there is no horizon and the metric can be thought 
to describe the gravitational field of a configuration of exotic matter (see 
Fig.~\ref{fig1}). The Bardeen metric can be seen as the prototype of a large class 
of non-Kerr BH metrics, in which the metric tensor in Boyer-Lindquist coordinates 
has the same expression of the Kerr one with $M$ replaced by a mass function 
$m(r)$ that depends only on the radial coordinate $r$ and that reduces to $M$ as 
$r \rar \infty$~\cite{bar1}. This class of metrics is particularly difficult to test, because 
the corresponding disk's thermal spectrum and iron line profile are extremely 
similar to the one corresponding to a Kerr BH with different spin~\cite{cb2}.

\begin{figure*}
\begin{center}
\hspace{-0.5cm}
\includegraphics[type=pdf,ext=.pdf,read=.pdf,width=8.5cm]{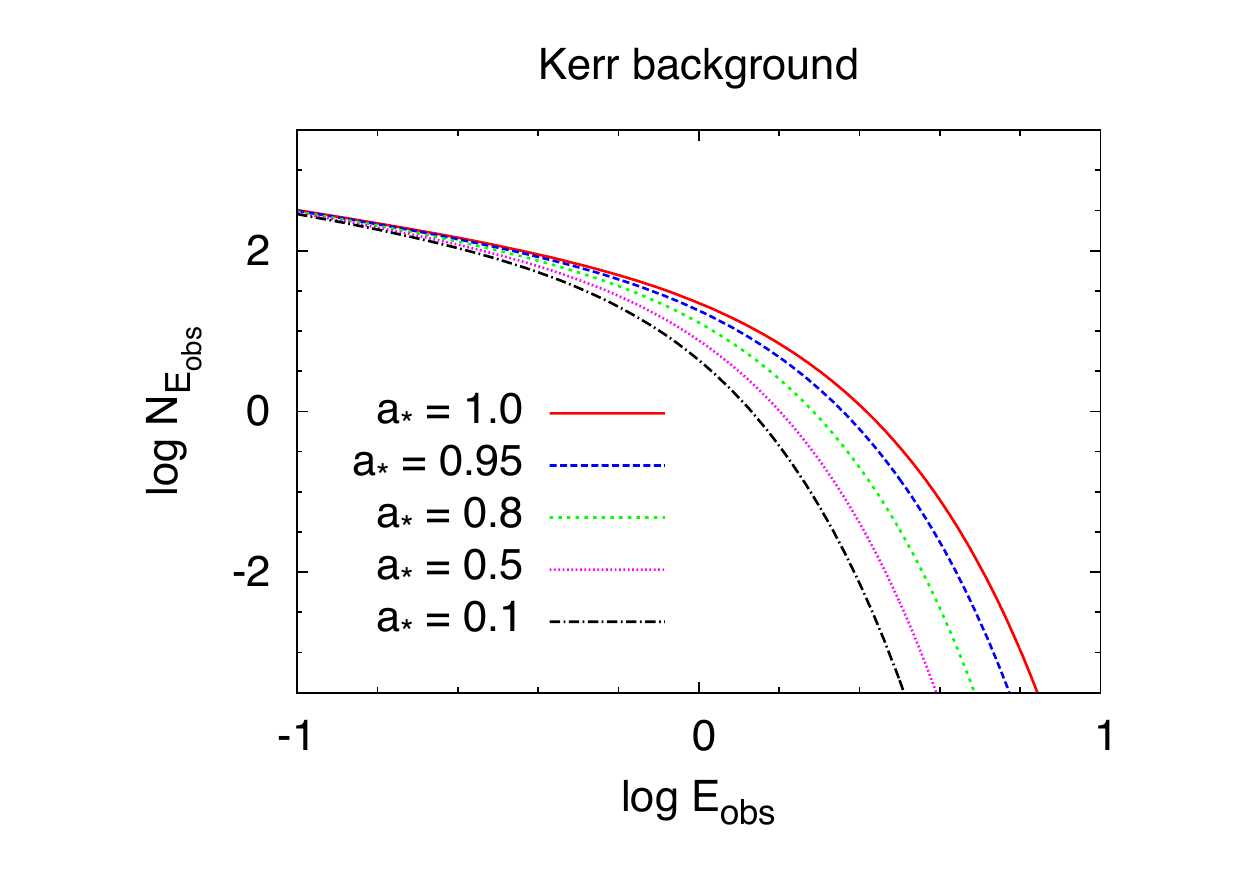}\\
\includegraphics[type=pdf,ext=.pdf,read=.pdf,width=8.5cm]{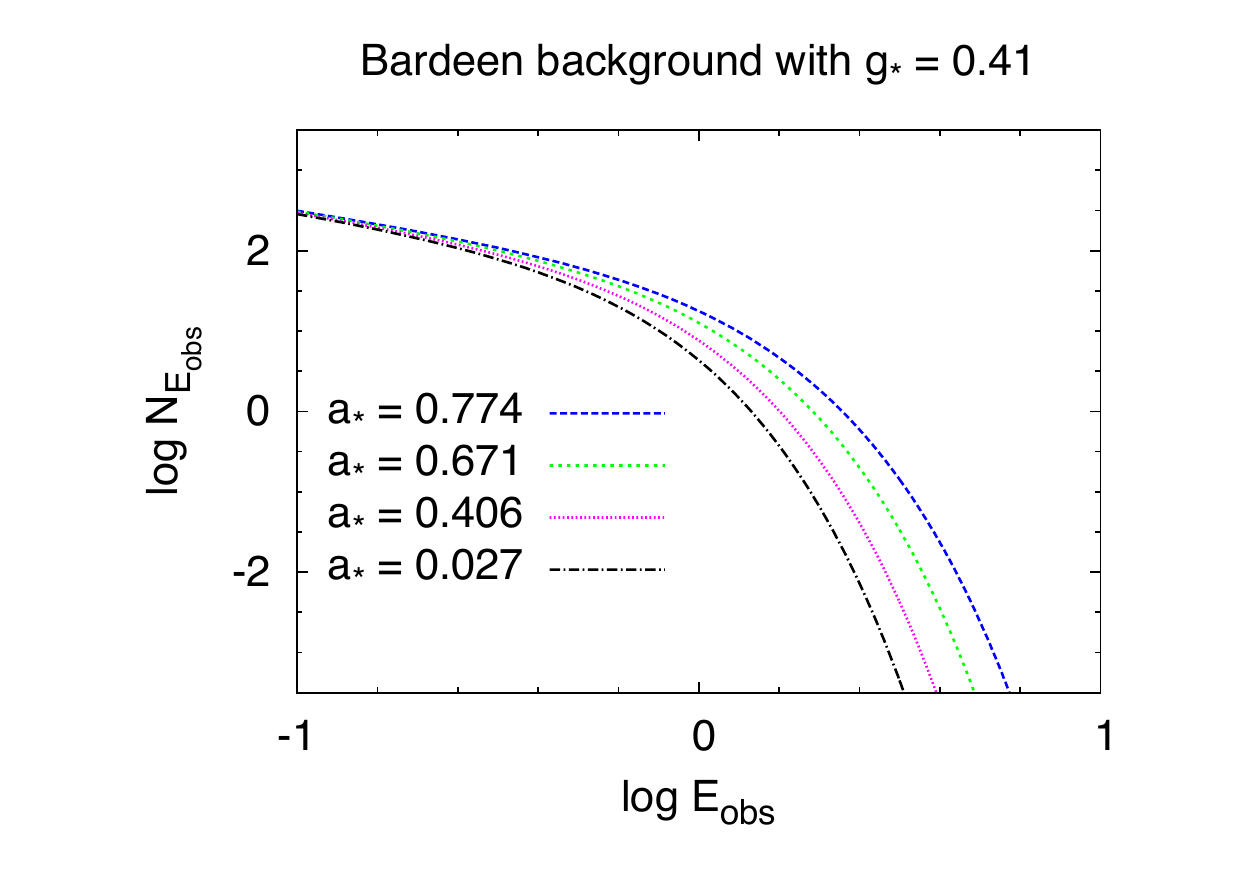}
\includegraphics[type=pdf,ext=.pdf,read=.pdf,width=8.5cm]{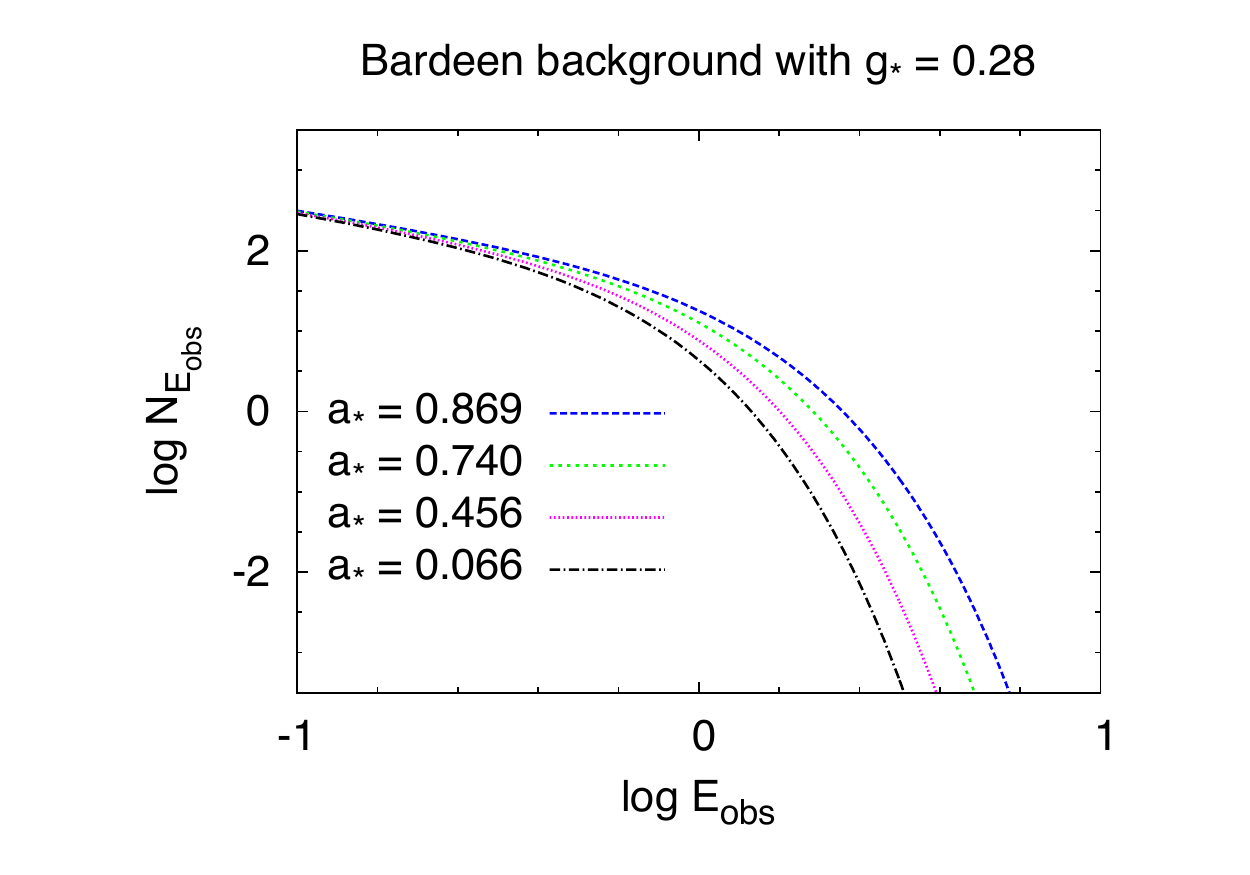}
\end{center}
\vspace{-0.5cm}
\caption{Thermal spectra of thin accretion disks around a Kerr BH (top panel),
a Bardeen BH with $g_* = 0.41$ (bottom left panel), a Bardeen BH with $g_* = 0.28$ 
(bottom right panel) for different values of the spin parameter $a_*$. The spectra 
on the bottom panels are extremely similar to the ones on the top panel with 
the same line style; that is, the spectrum of a Bardeen BH looks like the one of a 
Kerr BH with different spin. However, it is not always possible to mimic the spectrum 
of a very fast-rotating Kerr BH. For instance, a Bardeen BH with $g_* = 0.41$ cannot 
mimic a Kerr BH with $a_* = 1$, but only Kerr BHs with $a_* \lesssim 0.95$, because
the maximum value of the spin parameter of a Bardeen BH with $g_* = 0.41$ is 
$a_* \approx 0.775$. See the text for more details.}
\label{fig2}
\end{figure*}

The standard procedure to test the Bardeen metric with the X-ray data of the
BH candidate in Cygnus~X-1 would be to repeat the studies of Refs.~\cite{cyg,cyg2} with
the Kerr background replaced by the Bardeen one. This would require a detailed
analysis to include a large number of astrophysical effects and would be
very time consuming. However, we can arrive at the same result with a much 
faster approach. The key-point is that the thermal spectrum of a thin disk can be 
used to fit only one parameter of the background geometry. If we assume the Kerr 
metric, we can infer the BH spin parameter $a_*$. If we relax the Kerr BH 
hypothesis, we can measure some combination of the spin parameter and of 
possible deviations from the Kerr solution. As shown in Fig.~\ref{fig2}, the disk's
thermal spectrum of a Bardeen BH with specific values of $a_*$ and $g_*$
is practically indistinguishable from the one of a Kerr BH with spin $\tilde{a}_*$
($\neq a_*$). Bardeen BHs with the same $g_*$ and spin parameter higher
(lower) than $a_*$ look like Kerr BHs with spin parameter higher (lower) than
$\tilde{a}_*$. Since here the goal is to test the background metric around the 
BH candidate in Cygnus~X-1, we do not need to repeat the analysis of 
Refs.~\cite{cyg,cyg2} with the Bardeen background, but we can simply translate 
the Kerr measurements $a_* > 0.95$ and $a_* > 0.983$ to a bound on $a_*$
and $g_*$ by comparing the theoretical predictions of disk's thermal spectrum
around Kerr and Bardeen BHs.

{\it Calculation method ---}
The calculations of the thermal spectrum of a geometrically thin and optically 
thick accretion disk have been already extensively discussed in the literature, for 
both Kerr and non-Kerr spacetimes~\cite{cfm,cb}. The theoretical framework is the 
Novikov-Thorne model~\cite{nt}. The exact expression of the background metric 
enters the equations governing the time-averaged radial structure of the disk, the 
calculation of the propagation of the radiation from the disk to the distant observer, 
the motion of the particles of gas in the disk (determining the Doppler redshift and 
blueshift), and the inner edge of the disk, which is thought to be at the ISCO 
radius in the Novikov-Thorne model and eventually is the most important 
ingredient to infer the properties of the spacetime around the BH candidate.

The Novikov-Thorne model assumes that the disk is on the equatorial plane and 
that the disk's gas moves on nearly geodesic circular orbits. The time-averaged 
energy flux emitted from the surface of the disk is~\citep{nt}
\be
\mathcal{F}(r) = \frac{\dot{M}}{4 \pi M^2} F(r) \, ,
\ee
where $F(r)$ is the dimensionless function
\be
\hspace{-0.5cm}
F(r) = - \frac{\partial_r \Omega}{(E - \Omega L)^2} 
\frac{M^2}{\sqrt{-G}}
\int_{r_{\rm in}}^{r} (E - \Omega L) 
(\partial_\rho L) d\rho \, .
\ee
Here $E$, $L$, and $\Omega$ are, respectively, the conserved specific energy, 
the conserved axial-component of the specific angular momentum, and the 
angular velocity for equatorial circular geodesics; $G = - \alpha^2 g_{rr} g_{\phi\phi}$ 
is the determinant of the near equatorial plane metric, where $\alpha^2 =
g_{t\phi}^2/g_{\phi\phi} - g_{tt}$ is the lapse function; $r_{\rm in}$ is the inner 
radius of the accretion disk, which is assumed to be the radius of the ISCO. 
Since the disk is in thermal equilibrium, the emission is blackbody-like and 
we can define an effective temperature $T_{\rm eff} (r)$ from the relation 
$\mathcal{F}(r) = \sigma T^4_{\rm eff}$, where $\sigma$ is the Stefan-Boltzmann 
constant. The local specific intensity of the radiation emitted by the disk is
\be\label{eq-i-bb}
I_{\rm e}(\nu_{\rm e}) = \frac{2 h \nu^3_{\rm e}}{f_{\rm col}^4} 
\frac{\Upsilon}{\exp\left(\frac{h \nu_{\rm e}}{k_{\rm B} T_{\rm col}}\right) - 1} \, ,
\ee
where $T_{\rm col} (r) = f_{\rm col} T_{\rm eff}$ is the color temperature and
$f_{\rm col}$ is the color factor, which takes non-thermal effects into 
account ($f_{\rm col} \approx 1.6$ in our case). $\nu_{\rm e}$ is the photon 
frequency, $h$ is the Planck constant,
$k_{\rm B}$ is the Boltzmann constant, and $\Upsilon$ is a 
function of the angle between the wavevector of the photon emitted by the disk 
and the normal of the disk surface, say $\xi$. The two most common options are 
$\Upsilon = 1$ (isotropic emission) and $\Upsilon = \frac{1}{2} + \frac{3}{4} \cos\xi$ 
(limb-darkened emission).

The spectrum can be conveniently written in terms of the photon flux number 
density as measured by a distant observer:
\be\label{eq-n2}
\hspace{-0.5cm}
N_{E_{\rm obs}} &=&
\frac{1}{E_{\rm obs}} \int I_{\rm obs}(\nu) d \Omega_{\rm obs} = 
\nonumber\\ &=& 
\frac{1}{E_{\rm obs}} \int w^3 I_{\rm e}(\nu_{\rm e}) d \Omega_{\rm obs} = 
\nonumber\\ &=& 
A_1 \left(\frac{E_{\rm obs}}{\rm keV}\right)^2
\int \frac{1}{M^2} \frac{\Upsilon dXdY}{\exp\left[\frac{A_2}{g F^{1/4}} 
\left(\frac{E_{\rm obs}}{\rm keV}\right)\right] - 1} \, ,
\ee
where $I_{\rm obs}$, $E_{\rm obs}$, and $\nu$ are, respectively, the specific
intensity of the radiation, the photon energy, and the photon frequency measured
by the distant observer. $I_{\rm e}(\nu_{\rm e})/\nu_{\rm e}^3 = I_{\rm obs} 
(\nu_{\rm obs})/\nu^3$ follows from the Liouville theorem. $d\Omega_{\rm obs} 
= dX dY / D^2$ is the element of the solid angle subtended by the image of the 
disk on the observer's sky, $X$ and $Y$ are the coordinates of the position of 
the photon on the sky, as seen by the distant observer, while $D$ is the distance 
of the source. $A_1$ and $A_2$ are given by (reintroducing the constants 
$G_{\rm N}$ and $c$)
\be
A_1 &=&  
\frac{2 \left({\rm keV}\right)^2}{f_{\rm col}^4} 
\left(\frac{G_{\rm N} M}{c^3 h D}\right)^2 = 
\nonumber\\ &=& 
\frac{0.07205}{f_{\rm col}^4} 
\left(\frac{M}{M_\odot}\right)^2 
\left(\frac{\rm kpc}{D}\right)^2 \, 
{\rm \gamma \, keV^{-1} \, cm^{-2} \, s^{-1}} \, , \nonumber\\
A_2 &=&  
\left(\frac{\rm keV}{k_{\rm B} f_{\rm col}}\right) 
\left(\frac{G_{\rm N} M}{c^3}\right)^{1/2}
\left(\frac{4 \pi \sigma}{\dot{M}}\right)^{1/4} = 
\nonumber\\ &=& 
\frac{0.1331}{f_{\rm col}} 
\left(\frac{\rm 10^{18} \, g \, s^{-1}}{\dot{M}}\right)^{1/4}
\left(\frac{M}{M_\odot}\right)^{1/2} \, .
\ee
$w$ is the redshift factor and it turns out to be (for the details, see e.g.~\cite{cb})
\be\label{eq-red-g}
w = \frac{\sqrt{-g_{tt} - 2 g_{t\phi} \Omega - g_{\phi\phi} \Omega^2}}{1 + 
\lambda \Omega} \, ,
\ee
where $\lambda = k_\phi/k_t$ is a constant of the motion along the photon path.
Doppler boosting, gravitational redshift, and frame dragging are entirely encoded 
in the redshift factor $w$, while the effect of light bending enters the integral in
Eq.~\eqref{eq-n2}, as every photon on the sky is associated with its emission point
on the disk (this is done by integrating backward in time null geodesics from the 
observer to the disk).

In the Kerr spacetime, the model has five free parameters: the BH mass, $M$,
the mass accretion rate, $\dot{M}$, the spin parameter, $a_*$, the disk inclination
angle with respect to the line of sight of the distant observer, $i$, and the distance
of the object, $D$. If $M$, $i$, and $D$ can be estimated from independent
measurements (e.g. optical observations), one can fit the X-ray data of the thermal
spectrum of the disk and infer $a_*$ and $\dot{M}$\footnote{Actually, for Cygnus~X-1 
the situation is more complicated and one has to fit also other features of the spectrum.
However, the measurement of $a_*$ is obtained from the properties of the thermal 
spectrum of the accretion disk~\cite{cyg,cyg2}.}. In the Bardeen metric, there is a sixth
free parameter, $g_*$. However, as discussed above, the spectrum is somehow
degenerate with respect $a_*$ and $g_*$, in the sense that these two parameters 
cannot be determined independently, but it is only possible to infer a certain
combination of them. In principle, the mass accretion rate $\dot{M}$ can be 
determined independently, from the low-energy part of the spectrum, whose photons 
are mainly emitted at large radii, where gravity is almost Newtonian. This is not
really true in the case of X-ray data of stellar-mass BH candidates. However, in
the case of good measurement it is a good approximation to assume that the
determination of $a_*$ (or $a_*$ and $g_*$) is not correlated to the one of 
$\dot{M}$.

{\it Results ---}
We can now compare the thermal spectrum of thin disks around Kerr and
Bardeen BHs to translate the bound of the spin parameter of the BH candidate in
Cygnus~X-1 found in Refs.~\cite{cyg,cyg2} under the assumption of Kerr geometry 
into an allowed region on the spin parameter-deformation parameter plane
of the Bardeen solution. We can make the calculation with the physical parameter
of Cygnus~X-1 ($M = 14.8$~$M_\odot$, $i = 27.1$~deg, $D =1.86$~kpc, and 
even $\dot{M} = 1.3 \cdot 10^{17}$~g/s, which is roughly the mean mass accretion 
rate found from the data analyzed in~\cite{cyg,cyg2}). However, the final result is
quite independent of this choice, as in the end we compare Kerr and Bardeen
spectra with the same model parameters except the ones providing the geometry
of the spacetime. For more exotic metrics, this may not be true. For instance, 
a different Keplerian velocity and photon propagation may affect, respectively,
the Doppler effect and the light bending, so the final constraint may depend
on the disk's inclination angle $i$.

The 3$\sigma$-bound $a_* > 0.95$ found in~\cite{cyg} for the Kerr metric 
becomes the region inside the red solid line of Fig.~\ref{fig1} for the Bardeen  
background. The constraint is 
\be
a_* > 0.78 \, , \quad |g_*| < 0.41 \, .
\ee
The 3$\sigma$-bound $a_* > 0.983$ found in~\cite{cyg2} gives instead the 
region inside the blue dashed line of Fig.~\ref{fig1} and the bound is
\be
a_* > 0.89 \, , \quad |g_*| < 0.28 \, .
\ee
The region of objects without event horizon shown in Fig.~\ref{fig1} can be 
immediately ruled out for two reasons. First, like in the case of the Kerr metric
with $g_* = 0$~\cite{th10}, the thermal spectrum of disks around horizonless 
configurations is much harder, due to the sudden drop of the ISCO at smaller 
radii. Second, even if for $g_* \neq 0$ horizonless configurations can be 
created~\cite{bar2}, they are expected to be highly unstable, due to the 
existence of the ergoregion and of stable orbits with negative energy~\cite{ss}.

%%%%%%%%%%%%%%%%%%%%%%%%%%%%%%%

{\it Acknowledgments ---}
This work was supported by the NSFC grant No.~11305038, the Innovation 
Program of Shanghai Municipal Education Commission grant No.~14ZZ001, 
the Thousand Young Talents Program, and Fudan University.

%%%%%%%%%%%%%%%%%%%%%%%%%%%%%%

\end{document}